\documentclass[pra,aps,showpacs,twocolumn,twoside,superscriptaddress]{revtex4}
\usepackage{amssymb}

\usepackage{amsmath,amsfonts,amssymb,color,epsfig,graphics,graphicx,latexsym,revsymb,theorem,verbatim,bm}

\newtheorem{definition}{Definition}
\newtheorem{proposition}[definition]{Proposition}
\newtheorem{lemma}[definition]{Lemma}

\newtheorem{theorem}[definition]{Theorem}
\newtheorem{corollary}[definition]{Corollary}
\newtheorem{conjecture}[definition]{Conjecture}

\newtheorem{remark}[definition]{Remark}

\def\squareforqed{\hbox{\rlap{$\sqcap$}$\sqcup$}}
\def\qed{\ifmmode\squareforqed\else{\unskip\nobreak\hfil
\penalty50\hskip1em\null\nobreak\hfil\squareforqed
\parfillskip=0pt\finalhyphendemerits=0\endgraf}\fi}
\def\endenv{\ifmmode\;\else{\unskip\nobreak\hfil
\penalty50\hskip1em\null\nobreak\hfil\;
\parfillskip=0pt\finalhyphendemerits=0\endgraf}\fi}
\newenvironment{proof}{\noindent \textbf{{Proof.~} }}{\qed}

\def\bcj{\begin{conjecture}}
\def\ecj{\end{conjecture}}
\def\bcr{\begin{corollary}}
\def\ecr{\end{corollary}}
\def\bd{\begin{definition}}
\def\ed{\end{definition}}
\def\bea{\begin{eqnarray}}
\def\eea{\end{eqnarray}}
\def\be{\begin{equation}}
\def\ee{\end{equation}}
\def\bl{\begin{lemma}}
\def\el{\end{lemma}}
\def\bpf{\begin{proof}}
\def\epf{\end{proof}}
\def\bpp{\begin{proposition}}
\def\epp{\end{proposition}}
\def\br{\begin{remark}}
\def\er{\end{remark}}
\def\bt{\begin{theorem}}
\def\et{\end{theorem}}

\def\d{\delta}

\def\r{\rho}

\def\ph{\varphi}

\def\ps{\psi}

\def\L{\Lambda}

\def\Ph{\Phi}
\def\Ps{\Psi}

\def\max{\mathop{\rm max}}

\newcommand{\nc}{\newcommand}

\newcommand{\bra}[1]{\langle#1|}
\newcommand{\ket}[1]{|#1\rangle}
\newcommand{\proj}[1]{| #1\rangle\!\langle #1 |}

\newcommand{\braket}[2]{\langle#1|#2\rangle}

\newcommand{\abs}[1]{|#1|}

\newcommand{\Ra}{\Rightarrow}

\newcommand{\ox}{\otimes}

\nc{\cA}{{\cal A}} \nc{\cB}{{\cal B}} \nc{\cC}{{\cal C}}
\nc{\cD}{{\cal D}} \nc{\cE}{{\cal E}} \nc{\cF}{{\cal F}}
\nc{\cG}{{\cal G}} \nc{\cH}{{\cal H}} \nc{\cI}{{\cal I}}
\nc{\cJ}{{\cal J}} \nc{\cK}{{\cal K}} \nc{\cL}{{\cal L}}
\nc{\cM}{{\cal M}} \nc{\cN}{{\cal N}} \nc{\cO}{{\cal O}}
\nc{\cP}{{\cal P}} \nc{\cR}{{\cal R}} \nc{\cS}{{\cal S}}
\nc{\cT}{{\cal T}} \nc{\cX}{{\cal X}} \nc{\cZ}{{\cal Z}}

\begin{document}
\title{ Connections of geometric measure of entanglement of pure symmetric states to quantum state estimation}

\author{Lin Chen}
\affiliation{Centre for Quantum Technologies, National University of
Singapore, 3 Science Drive 2, Singapore 117543}
\email{cqtcl@nus.edu.sg (Corresponding~Author)}

\author{Huangjun Zhu}
\affiliation{Centre for Quantum Technologies, National University of
Singapore, 3 Science Drive 2, Singapore 117543} \affiliation{NUS
Graduate School for Integrative Sciences and Engineering, Singapore
117597, Singapore} \email{zhuhuangjun@nus.edu.sg}

\author{Tzu-Chieh Wei}
\affiliation{Department of Physics and Astronomy, University of British
Columbia, Vancouver, British Columbia V6T 1Z1, Canada}\email{twei@phas.ubc.ca}
\begin{abstract}
We study the geometric measure of entanglement (GM) of pure
symmetric states related to rank-one positive-operator-valued
measures (POVMs) and establish a general connection with quantum
state estimation theory, especially the maximum likelihood
principle. Based on this connection, we provide a method for
computing the GM of these states  and demonstrate its additivity
property under certain conditions. In particular, we prove the
additivity of the GM of pure symmetric multiqubit states whose
Majorana points under Majorana representation are distributed within
a half sphere, including all pure symmetric three-qubit states. We
then introduce a family of symmetric states that are generated from
mutually unbiased bases (MUBs), and derive an analytical formula for
their GM. These states include Dicke states as special cases, which
have already been realized in experiments. We also derive the GM of
symmetric states generated from symmetric informationally complete
POVMs (SIC~POVMs) and use it to characterize all inequivalent
SIC~POVMs in three-dimensional Hilbert space that are covariant with
respect to the Heisenberg--Weyl group. Finally, we describe an
experimental scheme for creating the symmetric multiqubit states
studied in this article and a possible scheme for measuring the
permanent of the related Gram matrix.
\end{abstract}

\date{ \today }



\maketitle

\section{\label{sec:introduction} introduction}
Quantum entanglement is a crucial resource for quantum computation
\cite{Joz97,RB01} and other  information processing tasks,
such as quantum teleportation~\cite{BBCJPW93}, superdense coding~\cite{BW92} and quantum key distribution \cite{Eke91}. 
In the past few decades, there have been tremendous efforts in understanding
various aspects of entanglement in both bipartite and multipartite settings.
One of the central issues in entanglement theory is the characterization and
quantification of multipartite entanglement \cite{PV07,HHHH09}. Among the many
approaches to the investigation of entanglement, several geometrically
motivated measures have been proposed and proved to be useful, such as
relative entropy of entanglement \cite{VPRK97,VP98}, \emph{geometric measure
of entanglement (GM)} \cite{Shi95,WG03,Bro01} and logarithmic global
robustness \cite{VT99,HN03}. Remarkably, these three measures turn out to be
related~\cite{HMM06,HMM08,WEGM04,Cav06,Wei08,ZCH10}.

Among the three measures, GM seems to be the easiest to handle and has thus
started to attract attention in recent years. Moreover, its applications have
gone beyond entanglement theory. GM is closely related to the construction of
optimal entanglement witnesses \cite{WG03}, experimental estimation of
entanglement~\cite{GRW07}, and discrimination of quantum states under local
operations and classical communications (LOCC) \cite{HMM06,HMM08,MMV07}.
Recently, GM has also been utilized to determine the universality of resource
states for one-way quantum computation \cite{NDMB07,MPMNDB10}, and to study
 generic multipartite pure states  as computational resources
\cite{GFE09,ZCH10}. In the context of condensed matter physics, GM has been
demonstrated to be useful for studying quantum many-body systems, such as
characterizing ground state properties and detecting phase transitions
\cite{WDMVG05,ODV08,Oru08,Oru08b,Wei10a,OW09,SAPFV10}.

Given the above applications of GM, it is thus interesting to
explore its connection with other important research fields, such as
quantum state estimation \cite{PR04,LR09}.  Quantum state estimation
is  a procedure of inferring the state of a quantum system by
general measurements---positive-operator-valued measures (POVMs). It
is a central issue of quantum mechanics and a cornerstone  of
various quantum-information processing tasks, such as quantum
computation, quantum communication and quantum cryptography.  In
this context, many measurement and reconstruction  methods have been
proposed to estimate the target state. Among the reconstruction
methods, an efficient one is the \emph{maximum likelihood method},
which has been widely used in experiments~\cite{Hra97,PR04,
RHJ01,RH07,JKMW01,dldg08,LR09,as10}.

Concerning measurement schemes, \emph{mutually unbiased bases
(MUBs)} \cite{Iva81,WF89,Woo04, DEBZ10, App08} and \emph{symmetric
informationally complete (SIC) POVMs} \cite{Zau99,Fuc02,RBSC04,
App05,ADF07, AFF09,SG10,Zhu10}, which stand for very efficient von
Neumann measurements and POVMs \cite{Sco06}, respectively, are two
focuses in the current research community.  Great efforts have been
directed to solving their existence problem and understanding their
structure
\cite{Iva81,WF89,Woo04,App08,Zau99,RBSC04,App05,AFF09,SG10,Zhu10};
since they are closely related to the physics in finite-dimensional
Hilbert spaces~\cite{DEBZ10,Fuc02,ADF07}.

In this article, we aim to establish a general connection between GM
and quantum state estimation theory, and MUBs, SIC~POVMs in
particular. The cross fertilizing of these geometric ideas may bring
insights to both research fields. We begin by studying the GM of
pure symmetric states related to rank-one POVMs, and establishing a
general connection with quantum state estimation theory, especially
the maximum likelihood principle \cite{Hra97,PR04,RHJ01,RH07,LR09}.
Based on this connection, we provide a method for computing the GM
of these states and demonstrate its additivity property under
certain conditions. In particular, we prove the additivity of the GM
of pure symmetric multiqubit states whose Majorana points  under
Majorana representation \cite{Maj32, BKMGLS09,MGBBB10, Mar10,AMM10}
are distributed within a half sphere, including all pure symmetric
three-qubit states. We then introduce a family of  symmetric states
that are generated from MUBs, and derive an analytical formula for
their GM. These states reduce to Dicke states in special cases,
which are useful for quantum communication and have been realized in
experiments \cite{KSTSW07,WKKMTW09,PCTPWKZ09}. Next, we compute the
GM of symmetric states generated from SIC~POVMs. This result is then
used to  characterize all inequivalent SIC~POVMs in
three-dimensional Hilbert space that are covariant with respect to
the Heisenberg--Weyl group \cite{Zau99, RBSC04, App05,Zhu10}.
Finally, we propose an experimental scheme for creating the
symmetric multiqubit states studied in this article.

The rest of the article is organized as follows. In
Sec.~\ref{sec:GenConnection}, we establish a general connection
between the GM of pure symmetric states and the maximum likelihood
principle in quantum state estimation theory. In
Sec.~\ref{sec:AddGM},  we prove the additivity of the GM of pure
symmetric multiqubit states whose Majorana points are distributed
within a half sphere, including all pure symmetric three-qubit
states. In Sec.~\ref{sec:MUBGM}, we introduce a family of symmetric
states generated from MUBs, and derive an analytical formula for
their GM. In Sec.~\ref{sec:SICGM}, we derive the GM of symmetric
states generated from SIC~POVMs and use it to characterize
inequivalent SIC~POVMs in three-dimensional Hilbert space. We
discuss experimental methods for realizing the symmetrized states of
this article in Sec.~\ref{sec:creation}. We conclude with a summary.

\section{\label{sec:GenConnection} GM of pure symmetric states: a connection with quantum state estimation}
In this section, we study the GM of pure symmetric states related to
rank-one POVMs, and  establish a general connection with quantum
state estimation theory, especially the maximum likelihood
principle. Based on this connection, the GM of many pure symmetric
states can be computed analytically, and its additivity property  be
demonstrated. In the next section, this connection will be used to
show the additivity of the GM of pure symmetric multiqubit states
whose Majorana points are distributed within a half sphere,
including all pure symmetric three-qubit states. Additional examples
related to MUBs and SIC~POVMs will be presented in the following
sections to illustrate this general idea.

\subsection{ Preliminary}

The geometric measure  of entanglement  measures  the maximum overlap between
a given state and the set of separable states, or equivalently, the set of
pure product states, and is defined as \cite{WG03,HMM06}
\begin{eqnarray}
  \Lambda^2(\rho)&:=&\max_{\sigma\in \mathrm{SEP}} \mathrm{tr}(\rho\sigma) =
\max_{|\Ph\rangle\in \mathrm{PRO}}    \label{eq:overlap}
  \langle\Ph|\rho|\Ph\rangle,\\
  G(\rho) &:=& -2~\log \Lambda(\rho).    \label{eq:GM}
\end{eqnarray}
Here ``SEP" denotes the set of separable states, and ``PRO" the set
of   pure product states that fully factorize; ``log" has base 2
throughout this article. Any pure product state maximizing
Eq.~(\ref{eq:overlap}) is a \emph{closest product state} of $\rho$.

For symmetric states, the computation of GM can be greatly
simplified due to a result in Refs.~\cite{HKW09,
ZCH10,HMMOV09,WS10}:
\begin{proposition}\label{pro:symmetric}
  The closest product state to any  pure or mixed symmetric
  $N$-partite state $\rho$ can be chosen to be symmetric; it is  necessarily symmetric if
  $N\geq3$:
   \bea
  \label{ea:puresymmetricGM}
  \L^2(\rho)
  &=& \max_{\ket{\ph_1}, \ldots, \ket{\ph_N}}
\bigg( \bigotimes^N_{j=1} \bra{\ph_j} \bigg)\rho \bigg(
\bigotimes^N_{j=1} \ket{\ph_j} \bigg)\nonumber\\
  &=& \max_{\ket{\ph}}
 \bra{\ph}^{\ox N}\rho  \ket{\ph}^{\ox N}.
\eea
\end{proposition}
Recently, the GM of pure symmetric  three-qubit states have been
derived based on this observation \cite{CXZ10}, and a class of
maximally entangled three-qubit states has also been
obtained~\cite{TWP09,CXZ10}. The problem is still open for more
general situations, although progress has been made with respect to
the connection between the singular values of a hypermatrix and the
GM~\cite{HS10}. Below, we shall provide many examples where
analytical solutions can be obtained.

Another key ingredient in our investigation is the maximum
likelihood (ML) principle of quantum state estimation
\cite{Hra97,PR04,RHJ01,RH07,LR09}. Consider state estimation using a
rank-one POVM composed of $M$ outcomes that are represented by
subnormalized pure projectors $\Pi_j=|\psi_j\rangle\langle\psi_j|$
such that $\sum_j\Pi_j=I$. If we are given $N$ copies of an unknown
input state and perform $N$ measurements independently, then the
outcome statistics  obey a multinomial distribution. Suppose outcome
$j$ occurs $n_j$ times for $j=1,2\ldots,M$ ($\sum_j n_j=N$); then
the frequency of obtaining outcome $j$ is $f_j=n_j/N$. In the
standard state reconstruction, the estimator is obtained by solving
the following set of equations:
\begin{eqnarray}\label{eq:standard}
\langle\psi_j|\rho|\psi_j\rangle=f_j \quad\forall j.
\end{eqnarray}
However, such a solution does not always exist.  The ML principle
consists in choosing a state $\r_{\mathrm{ML}}$ that maximizes the
likelihood functional $\mathcal{L}(\rho)$ as an estimator of the
true state \cite{PR04,LR09,Hra97,RHJ01,RH07},
\begin{eqnarray}\label{eq:likelihood}
\mathcal{L}(\rho)=\prod_{j=1}^M p_j^{n_j},
\end{eqnarray}
where $p_j=\langle\psi_j|\rho|\psi_j\rangle$ is the probability of
obtaining outcome $j$ given the input state $\rho$. If there exists
a state $\rho_{\mathrm{s}}$ that satisfies Eq.~(\ref{eq:standard}),
that is, the probabilities $p_j$ derived from this state coincide
with the frequencies $f_j$, then $\rho_{\mathrm{s}}$ is also the
maximum point of the likelihood functional. In general, there is an
efficient iterative algorithm for finding the ML estimator if the
POVM is informationally complete (IC)
\cite{PR04,LR09,Hra97,RHJ01,RH07}.  A POVM is IC if we can
reconstruct any input state according to the measurement statistics.
If the POVM is not IC,   the maximum of the likelihood functional
can still be computed efficiently, but the ML estimators are
generally not unique. In addition, the ML principle is also
applicable when the ($|\psi_j\rangle\langle\psi_j|$)'s form an
incomplete POVM, that is $\sum_j|\psi_j\rangle\langle\psi_j|=\Pi\leq
I$ \cite{PR04,LR09,Hra97,RHJ01,RH07} (be sure to distinguish
``complete" and ``informationally complete").

\subsection{ Connection}

We are now ready to show the connection between the GM of pure
symmetric states and quantum state estimation theory. Following the
above notation, assume $\sum_j|\psi_j\rangle\langle\psi_j|=\Pi\leq
I$, and the largest eigenvalue of $\Pi$ is 1. Define
$\Psi(\{n_j\})\rangle$ as the symmetrized state of the product state
$\bigotimes_{j=1}^M|\psi_j\rangle^{\otimes n_j}$,
\begin{eqnarray}
  \label{ea:symmetrizedstate}
  |\Psi(\{n_j\})\rangle=c P_\mathrm{sym}
  \bigotimes_{j=1}^M|\psi_j\rangle^{\otimes n_j},
  \end{eqnarray}
where  $P_\mathrm{sym}$ is the projector onto the symmetric
subspace, and $c$ is a normalization constant, which can be assumed
to be positive without loss of generality. The effect of the
projector $P_\mathrm{sym}$ is determined by its action on pure
product states. The action of  $P_\mathrm{sym}$ on the tensor
product of $N$ single-particle kets $|a_k\rangle$ is given by
\begin{eqnarray}
P_\mathrm{sym} \bigotimes_{k=1}^N
|a_k\rangle=\frac{1}{N!}\sum_{\sigma\in S_N} \bigotimes_{k=1}^N
|a_{\sigma(k)}\rangle,
\end{eqnarray}
where $S_N$ is the symmetry group of $N$ letters. Now suppose
$|a_k\rangle$ is the $k$-th member of the multiset consisting of
$n_1$ copies of $|\psi_1\rangle$, $n_2$ copies of $|\psi_2\rangle$
and so on. Define $A$ as the Gram matrix of the kets $|a_k\rangle$,
i.e.,
\begin{eqnarray}
A_{jk}=\langle a_j|a_k\rangle, \quad j,k=1,2,\ldots,N.
\end{eqnarray}
The dependence of $A$ on $|\psi_j\rangle$ and $n_j$ has been
suppressed to simplify the notation. The normalization constant $c$
is thus given by $c=\sqrt{N!/\mathrm{perm}(A)}$~\cite{Wei10b}, where
$\mathrm{perm}(A)$ denotes the permanent of the matrix $A$ (for
other connections to permanent, see Ref.~\cite{WS10}).

According to Proposition~\ref{pro:symmetric}, the GM of
$|\Psi(\{n_{j}\})\rangle$ reads
\begin{eqnarray}
 \Lambda^2(|\Psi(\{n_{j}\})\rangle)&=&\max_{|\varphi\rangle}
 |\langle\varphi|^{\otimes N}|\Psi(\{n_j\})\rangle|^2\nonumber\\
 &=&\frac{N!}{\mathrm{perm}(A)}\max_{|\varphi\rangle}\prod_{j=1}^M|\langle\varphi|\psi_j\rangle|^{2n_{j}}\nonumber\\
 &\le&\frac{N!}{\mathrm{perm}(A)}\max_{\r}\cL(\r)\nonumber\\
 &\leq&\frac{N!}{\mathrm{perm}(A)}\max_{ 0\leq p_{j}\leq 1,\; \sum_{j}p_{j}=1}\;\prod_{j=1}^M p_{j}^{n_{j}}\nonumber\\
 &=&\frac{N!}{\mathrm{perm}(A)}\prod_{j=1}^M f_{j}^{n_{j}}
 \label{ea:upperboundforlikelihood}.
\end{eqnarray}
Note that, apart from a constant factor, the functional to be
maximized in the second line of the above equation is exactly the
likelihood functional $\mathcal{L}(\rho)$ associated with the POVM
$\{\Pi_j\}$ if we take $f_j$ as ``frequency" [see
Eq.~(\ref{eq:likelihood})], except that $\rho$ is restricted to pure
states here; this is the reason for the first inequality in
Eq.~(\ref{ea:upperboundforlikelihood}). The ``frequencies" $f_j$ are
called \emph{compatible} if there exists a normalized pure state
$|\varphi_{\mathrm{s}}\rangle$ such that
$|\langle\varphi_{\mathrm{s}}|\psi_j\rangle|^2=f_{j}\;\forall j$;
that is, the $f_j$'s  can coincide with the probabilities derived
from a pure state. Note that such a pure state  is necessarily
contained in the eigenspace to the largest eigenvalue of $\Pi$. The
maximum in the last line of Eq.~(\ref{ea:upperboundforlikelihood})
can be obtained if and only if the $f_j$'s are compatible.  In that
case, the likelihood functional $\cL(\r)$ is maximized at the pure
state $|\varphi_{\mathrm{s}}\rangle$, and
$|\varphi_\mathrm{s}\rangle^{\otimes N}$ is a closest product state
to $|\Psi(\{n_{j}\})\rangle$.

Moreover, if the maximum of $\mathcal{L}(\rho)$ can be obtained at a
pure state, which is true if the $f_j$'s are compatible, then the GM
of $|\Psi(\{n_{j}\})\rangle$ and any pure symmetric $N$-partite
state $|\Phi\rangle$ is additive; that is,
$G(|\Psi(\{n_{j}\})\rangle \ox
\ket{\Ph})=G(|\Psi(\{n_{j}\})\rangle)+G(\ket{\Ph})$ (see
Ref.~\cite{ZCH10} for a detailed discussion on the additivity
property of GM). We prove this statement in Appendix A. In that
case, the GM of $|\Psi(\{n_{j}\})\rangle$ is equal to its asymptotic
regularized quantity and gives a lower bound for asymptotic relative
entropy of entanglement and asymptotic logarithmic global robustness
\cite{ZCH10}. For convenience,  we summarize the above observations
as the following theorem.

\begin{theorem}\label{thm:POVMstateGM}
Suppose  there are $M$ kets $|\psi_j\rangle$  (not necessarily
normalized) that satisfy
$\sum_{j=1}^M|\psi_j\rangle\langle\psi_j|=\Pi$, and the largest
eigenvalue of $\Pi$ is 1; $n_1, n_2,\ldots, n_M$ are non-negative
integers with sum $N=\sum_j n_j$ and $f_j=n_j/N$. Define $A$ as the
Gram matrix of the multiset consisting of $n_1$ copies of
$|\psi_1\rangle$, $n_2$ copies of $|\psi_2\rangle$ and so on; and
$|\Psi(\{n_j\})\rangle=
\sqrt{N!/\mathrm{perm}(A)}P_{\mathrm{sym}}\bigotimes_{j=1}^M|\psi_j\rangle^{\otimes
n_j}$. Then the  GM of  $|\Psi(\{n_j\})\rangle$ is lower bounded by
\begin{eqnarray}\label{eq:POVMstateGM}
G(|\Psi(\{n_{j}\})\rangle) &\geq
&-\log\biggl(\frac{N!}{\mathrm{perm}(A)}\prod_{j=1}^M
f_{j}^{n_{j}}\biggr);
\end{eqnarray}
the bound is saturated if and only if  the $f_j$'s are compatible,
that is, there exists a normalized pure state
$|\varphi_{\mathrm{s}}\rangle$ such that
\begin{eqnarray}\label{eq:compatible}
|\langle\varphi_{\mathrm{s}}|\psi_j\rangle|^2=f_{j}\quad\forall j.
\end{eqnarray}
Moreover, if the maximum of $\mathcal{L}(\rho)$ [see
Eq.~(\ref{eq:likelihood}) and (\ref{ea:upperboundforlikelihood})]
can be obtained at a pure state, which is true if the $f_j$'s are
compatible, then the GM of $|\Psi(\{n_{j}\})\rangle$ and  any pure
symmetric $N$-partite state $|\Phi\rangle$ is  additive; that is,
\begin{eqnarray}\label{eq:POVMstateGMadd}
G(|\Psi(\{n_{j}\})\rangle \ox
\ket{\Ph})=G(|\Psi(\{n_{j}\})\rangle)+G(\ket{\Ph}).
\end{eqnarray}
\end{theorem}

No matter whether Eq.~(\ref{eq:compatible}) can be satisfied or not,
the maximum of the likelihood functional $\cL(\r)$ in
Eq.~(\ref{ea:upperboundforlikelihood}) can be computed efficiently
with an iterative algorithm \cite{Hra97,PR04,RHJ01,RH07,LR09}. It
gives an upper bound for $\Lambda^2(|\Psi(\{n_j\})\rangle)$ and thus
a lower bound for $G(|\Psi(\{n_j\})\rangle)$. On the other hand,
when Eq.~(\ref{eq:compatible}) is not satisfied, it is still
possible that the maximum of the likelihood functional is obtainable
at a pure state. Then the $N$-fold tensor product of the pure state
is a closest product state to $|\Psi(\{n_j\})\rangle$, and the GM of
$|\Psi(\{n_{j}\})\rangle$ and any pure symmetric $N$-partite  state
is additive (see Sec.~\ref{sec:AddGM} for examples). In other words,
the compatibility condition on the $f_j$'s is sufficient but not
necessary for Eq.~(\ref{eq:POVMstateGMadd}) to hold. It is really
remarkable that we can derive the GM of these symmetric states and
its additivity property from the property of the likelihood
functional.

Besides, the  condition on the largest eigenvalue of $\Pi$ in
Theorem~\ref{thm:POVMstateGM} is not essential; it is adopted mainly
for a closer connection with state estimation theory. If instead the
largest eigenvalue of $\Pi$  is $g>0$, then
Theorem~\ref{thm:POVMstateGM} is still applicable as long as $f_j$
is replaced by $gf_j$ in Eqs.~(\ref{eq:POVMstateGM}) and
(\ref{eq:compatible}). One advantage of this alternative convention
is that $|\psi_j\rangle$ can now be chosen to be normalized. A
simple example in the case of qubits is as follows. Suppose there
are $M$ normalized kets $|\psi_j\rangle$  whose Bloch vectors are
distributed on a circular cone around the $z$ axis, such that
$\sum_{j=1}^M|\psi_j\rangle\langle\psi_j|=M(I+r \sigma_z)/2$ with
$0< r <1$; and
$|\Psi\rangle=\sqrt{M!/\mathrm{perm}(A)}P_{\mathrm{sym}}\bigotimes_{j=1}^M|\psi_j\rangle$.
Then $|\Psi\rangle$ has a unique closest product state
$|0\rangle^{\otimes M}$ and its  GM  is
\begin{eqnarray}
G(|\Psi\rangle)
&=&-\log\biggl[\frac{M!}{\mathrm{perm}(A)}\Bigl(\frac{1+r}{2}\Bigr)^M\biggr].
\end{eqnarray}

\section{\label{sec:AddGM} Additivity of GM of pure symmetric three-qubit states}
Every pure symmetric $N$-qubit state $|\Psi\rangle$ can be written
in the form in Eq.~(\ref{ea:symmetrizedstate}), that is,
\begin{eqnarray}
|\Psi\rangle\propto
P_{\mathrm{sym}}\bigotimes_{j=1}^M|\psi_j\rangle^{\otimes n_j};
\end{eqnarray}
and this representation is  unique up to permutations of the
$|\psi_j\rangle$'s and some phase factors. This  is well-known as
the Majorana representation \cite{Maj32}. Under this representation,
each pure symmetric $N$-qubit state corresponds to $N$ points on the
Bloch sphere.  Following Refs.~\cite{Mar10, AMM10}, the
$|\psi_j\rangle$'s are called Majorana points  of $|\Psi\rangle$.
Recently, Majorana representation has found many applications in the
study of multipartite entanglement, such as classification of
entanglement of pure symmetric multiqubit states under stochastic
LOCC (SLOCC) \cite{BKMGLS09}, investigation  of the GM of these
states and maximally entangled states among them
\cite{Mar10,MGBBB10,AMM10}.


As  an important application of the theory developed  in the
previous section and the Majorana representation, in this section,
we prove the additivity of the GM of pure symmetric multiqubit
states whose Majorana points are distributed within a half sphere,
which is true for all pure symmetric three-qubit states. As we shall
see shortly, the structure of the state space and its boundary plays
a crucial role in proving this additivity property.
\begin{theorem}\label{thm:SymMulQubit}
Suppose $|\Psi\rangle$ is a pure symmetric $N$-qubit state whose
Majorana points are distributed within a half sphere under the
Majorana representation; then the GM of $|\Psi\rangle$ and any pure
symmetric $N$-partite state is additive.
\end{theorem}
According to Theorem~\ref{thm:POVMstateGM}, to prove this theorem,
it suffices to show that the maximum of the following functional can
be obtained at a pure state,
\begin{eqnarray}
\mathcal{L}(\rho)=\prod_{j=1}^M
(\langle\psi_j|\rho|\psi_j\rangle)^{n_j}.
\end{eqnarray}
Without loss of generality, we can assume that the Majorana points
of $|\Psi\rangle$ lie within a half sphere with $z\geq0$, and the
Bloch vector of $\rho$ is $\bm{r}=(x,y,z)$ with $x^2+y^2+z^2\leq 1$.
Then it is straightforward to verify that $\mathcal{L}(\rho)$ is
nondecreasing with $z$, and thus its maximum can be obtained at the
boundary of the Bloch sphere, that is, at a pure state. This
completes the proof of Theorem~\ref{thm:SymMulQubit}.

For any pure symmetric three-qubit state, the three Majorana points
 lie within some half sphere;  hence, Theorem~\ref{thm:SymMulQubit} is applicable.
 The same is true for any
pure symmetric multiqubit state that has at most three distinct
Majorana points.
\begin{corollary}\label{cor:sym3Qubit}
The GM of any pure symmetric $N$-qubit  state that has at most three
distinct Majorana points and  any pure symmetric $N$-partite state
is additive. In particular, the GM of any pure symmetric three-qubit
state and any pure symmetric tripartite state is additive .
\end{corollary}

\begin{corollary}\label{cor:sym3points}
Suppose $|\Psi\rangle$ is a pure symmetric $N$-partite state which
can be written in the form
\begin{eqnarray}
|\Psi\rangle\propto
P_{\mathrm{sym}}\bigotimes_{j=1}^3|\psi_j\rangle^{\otimes n_j},
\end{eqnarray}
where $n_1+n_2+n_3=N$; then the GM of $|\Psi\rangle$ and any pure
symmetric $N$-partite  state is additive.
\end{corollary}
Without loss of generality, we can assume the three states
$|\psi_j\rangle$ belong to a three-dimensional Hilbert space. As in
the case of qubit, they  can be seen as three extremal points of the
eight-dimensional state space whose origin is the completely mixed
state; however, the boundary of the state space is no longer a
sphere, and the states on the boundary are not necessarily pure. In
addition, we can find  a suitable hyperplane passing through the
origin such that the three points are on the same side of the
hyperplane (or on the hyperplane). According to a similar reasoning
that leads to Theorem~\ref{thm:SymMulQubit} and
Corollary~\ref{cor:sym3Qubit}, the maximum of the following
functional
\begin{eqnarray}
\mathcal{L}(\rho)=\prod_{j=1}^3
(\langle\psi_j|\rho|\psi_j\rangle)^{n_j}
\end{eqnarray}
can be obtained at a state $\rho_{\mathrm{ML}}$ on the boundary of
the state space, whose rank is at most two. If $\rho_{\mathrm{ML}}$
is pure, then we are done. Otherwise, when $\rho$ is restricted to
the support of $\rho_{\mathrm{ML}}$, we have
\begin{eqnarray}
\mathcal{L}(\rho)=\prod_{j=1}^3
(\langle\psi_j^\prime|\rho|\psi_j^\prime\rangle)^{n_j}.
\end{eqnarray}
where $|\psi_j^\prime\rangle$ is the projection of $|\psi_j\rangle$
onto the support of $\rho_{\mathrm{ML}}$. Now applying the same
reasoning that leads to Corollary~\ref{cor:sym3Qubit} shows that the
maximum of $\mathcal{L}(\rho)$ can be obtained at a pure state.
Therefore, the Corollary follows from Theorem~\ref{thm:POVMstateGM}.

Theorem~\ref{thm:SymMulQubit} and Corollaries~\ref{cor:sym3Qubit}
and \ref{cor:sym3points} provide a method for computing the
asymptotic GM, which in turn provides a lower bound for the
asymptotic relative entropy of entanglement and the asymptotic
logarithmic global robustness \cite{ZCH10}. They are also useful in
the  study of multipartite pure states  as computational resources,
since  GM and its additivity property  are closely related to
whether these states are universal for quantum computation
\cite{NDMB07, MPMNDB10,GFE09,ZCH10}. Corollary~\ref{cor:sym3Qubit}
also implies the multiplicativity of the output purity of the
quantum channels associated with pure symmetric three-qubit states
according to the Werner-Holevo recipe \cite{WH02,ZCH10}.

\section{\label{sec:MUBGM} symmetric states generated from mutually unbiased bases}

To illustrate the general idea presented in
Sec.~\ref{sec:GenConnection}, in this section we consider the
situation where the POVM can be decomposed into a set of von Neumann
measurements, in particular the scenario where the  bases of the von
Neumann measurements are mutually unbiased \cite{DEBZ10}. We first
consider pure symmetric states generated from two bases of the qubit
Hilbert space, which reduce to Dicke states as special cases. We
then generalize the idea to higher-dimensional Hilbert spaces and
point out the role played by mutually unbiasedness.

\subsection{\label{sec:gDickeGM} Generalization of Dicke states}

Given $\bm{r}_0=(0,0,1)$ and  $\bm{r}_1=(\sin\theta,0,\cos\theta)$,
let $|0\rangle$, $|1\rangle$ denote the eigenbasis of
$\bm{r}_0\cdot\bm{\sigma}=\sigma_z$, and
$|\theta_+\rangle=\cos(\theta/2)|0\rangle+\sin(\theta/2)|1\rangle$,
$|\theta_-\rangle=\sin(\theta/2)|0\rangle-\cos(\theta/2)|1\rangle$
the eigenbasis of $\bm{r}_1\cdot\bm{\sigma}$.  Given four
nonnegative integers $\{n_{jk}\}=\{n_{00},n_{01};n_{10},n_{11}\}$,
let $N=n_{00}+n_{01}+n_{10}+n_{11}$,  $N_j=n_{j0}+n_{j1}$,
$f_{jk}=n_{jk}/N_j$ (assuming $N_j\neq 0$). Define $A$ as the Gram
matrix of the multiset consisting of  $n_{00}, n_{01}, n_{10},
n_{11}$ copies of $|0\rangle, |1\rangle, |\theta_+\rangle,
|\theta_-\rangle$, respectively; and
\begin{eqnarray}\label{eq:gDicke}
\bigl|\Psi\bigl(\theta,\{n_{jk}\}\bigr)\bigr\rangle:&=&\sqrt{\frac{N!}{\mathrm{perm}(A)}}P_\mathrm{sym}
\Bigl(|0\rangle^{\otimes n_{00}}\otimes|1\rangle^{\otimes
n_{01}}\nonumber\\
&&\otimes|\theta_+\rangle^{\otimes
n_{10}}\otimes|\theta_-\rangle^{\otimes n_{11}}\Bigr).
\end{eqnarray}
Here  the permanent $\mathrm{perm}(A)$ can be computed efficiently,
see Appendix B. Note that $|\Psi(\theta,\{n_{jk} \})\rangle$ can be
seen as a generalization of the Dicke state; it reduces to the Dicke
state when $\theta=0,\pi$ or $N_0=0$ or $N_1=0$. The GM of the Dicke
state has been derived in Ref.~\cite{WG03}, and its additivity
property has been demonstrated in Ref.~\cite{ZCH10}. When
$N_0,N_1\neq 0$, we have the following theorem.
 \begin{theorem}
 \label{thm:DickeGM}The GM of $|\Psi(\theta,\{n_{jk}\})\rangle$ is
lower bounded by
\begin{eqnarray}\label{eq:gDickeGM}
&&G(|\Psi(\theta,\{n_{jk}\})\rangle)
\geq-\mathrm{log}\Bigl(\frac{N!}{\mathrm{perm}(A)}f_{00}^{n_{00}}f_{01}^{n_{01}}f_{10}^{n_{10}}f_{11}^{n_{11}}\Bigr);\nonumber\\
\end{eqnarray}
the bound is saturated  if and only if there exists a qubit state
$|\varphi\rangle$ such that
\begin{eqnarray}\label{eq:CPsConstraint2}
|\langle\varphi|0\rangle|^2=f_{00},
\quad|\langle\varphi|1\rangle|^2=f_{01},\nonumber\\
|\langle\varphi|\theta_+\rangle|^2=f_{10},\quad
|\langle\varphi|\theta_-\rangle|^2=f_{11}.
\end{eqnarray}
When $\theta\neq0,\pi$, this condition  is equivalent to the
following one:
\begin{eqnarray}\label{eq:constraint}
|h_0\bm{s}_0+h_1\bm{s}_1|\leq 1,
\end{eqnarray}
where $h_j=f_{j0}-f_{j1}$, and $\bm{s}_0=(-\cot\theta,0,1)$,
$\bm{s}_1=(\csc\theta,0,0)$ is the dual basis of $\bm{r}_0,\bm{r}_1$
in the $x$-$z$ plane.

The GM of $|\Psi(\theta,\{n_{jk}\})\rangle$  and any pure symmetric
$N$-partite state is  additive, irrespective whether the condition
Eq.~(\ref{eq:CPsConstraint2}) is satisfied or not.
 \end{theorem}

We can briefly show Theorem~\ref{thm:DickeGM} as follows. First
Eqs.~(\ref{eq:gDickeGM}) and (\ref{eq:CPsConstraint2}) can be
derived according to a similar reasoning that leads to
Theorem~\ref{thm:POVMstateGM}.  Equation~(\ref{eq:constraint}) has a
nice geometric interpretation under the Bloch sphere representation.
Suppose the Bloch vector of $|\ph\rangle$ is $\bm{s}$. The first two
equations in Eq.~(\ref{eq:CPsConstraint2}) restrict $\bm{s}$ to a
plane satisfying $\bm{s}\cdot\bm{r}_0=h_0$, which is perpendicular
to $\bm{r}_0$, and the last two restrict $\bm{s}$ to a plane
satisfying $\bm{s}\cdot\bm{r}_1=h_1$, which is perpendicular to
$\bm{r}_1$. There exists a pure state satisfying
Eq.~(\ref{eq:CPsConstraint2}) if and only if the intersection of the
two planes passes through the Bloch sphere; that is, the
intersection of the two planes and the $x$-$z$ plane lies within the
unit circle centered at the origin on the $x$-$z$ plane. This is
exactly what  Eq.~(\ref{eq:constraint}) means. In the special case
$\theta=\frac{\pi}{2}, \frac{3\pi}{2}$, where the two bases
$|0\rangle,|1\rangle$ and $|\theta_\pm\rangle=|\pm\rangle$ are
mutually unbiased, Eq.~(\ref{eq:constraint}) simplifies to
$h_0^2+h_1^2\leq1$.  The additivity property of the GM of
$|\Psi(\theta,\{n_{jk}\})\rangle$ follows from
Theorem~\ref{thm:SymMulQubit}. This completes the proof of
Theorem~\ref{thm:DickeGM}.

When $n_{01}=n_{00}$ and $n_{11}=n_{10}$,
Eq.~(\ref{eq:CPsConstraint2}) is trivial to satisfy; and  the
$N$-fold tensor product of the two eigenstates of $\sigma_y$,
respectively, are  exactly two closest product states to
$|\Psi\bigl(\theta,\{n_{jk}\}\bigr)\bigr\rangle$. Besides,
Eq.~(\ref{eq:gDickeGM}) reduces to
\begin{eqnarray}
G(|\Psi(\theta,\{n_{jk}\})\rangle)&=&-\log\Bigl(\frac{N!}{2^N\mathrm{perm}(A)}\Bigr).
\end{eqnarray}
Remarkably, the GM is completely determined by the number of qubits
$N$ and $\mathrm{perm}(A)$. An interesting example is the state
$\bigl|\Psi\bigl(\theta,\{1,1;1,1\}\bigr)\bigr\rangle$ obtained when
$n_{00}=n_{01}=n_{10}=n_{11}=1$. It is a balanced four-qubit Dicke
state when $\theta=0,\pi$, and is equivalent to four-qubit GHZ state
under a suitable local unitary transformation when
$\theta=\frac{\pi}{2}, \frac{3\pi}{2}$, according to
Refs.~\cite{BKMGLS09,MGBBB10, Mar10,AMM10,Bas09}. Calculation shows
that $\mathrm{perm}(A)=[7+\cos(2\theta)]/2$, hence,
\begin{eqnarray}
G(|\Psi(\theta,\{1,1;1,1\})\rangle)&=&\log\frac{7+\cos(2\theta)}{3}.
\end{eqnarray}
The maximum is obtained at $\theta=0,\pi$, where the two bases
coincide, and the minimum at $\theta=\frac{\pi}{2},\frac{3\pi}{2}$,
where the two bases are mutually unbiased.

\subsection{A connection with MUBs}

To generalize the above idea to higher dimension, let
$|e^j_k\rangle$ for $j=0,1,\ldots,b-1, k=0,1,\ldots,d-1$ be $bd$
single-particle states such that the $d$ states for given $j$ form
an orthonormal basis \cite{DEBZ10}. Let $n_{jk}$ be a $b\times d$
matrix composed of nonnegative integers, $N=\sum_{j,k} n_{jk}$,
$N_j=\sum_{k}n_{jk}$, and $f_{jk}=n_{jk}/N_j$ (assuming $N_j\neq
0$). Define
\begin{eqnarray*}
\bigl|\Psi\bigl(\{n_{jk}\}\bigr)\bigr\rangle:=\sqrt{\frac{N!}{\mathrm{perm}(A)}}P_\mathrm{sym}
\bigotimes_{j,k} |e^j_k\rangle^{\otimes n_{jk}}.
\end{eqnarray*}
This state  reduces to a generalized Dicke state when there is only
one basis, that is $b=1$. According to the same reasoning as before,
the GM of $\bigl|\Psi\bigl(\{n_{jk}\}\bigr)\bigr\rangle$ is lower
bounded by
\begin{eqnarray*}
G(|\Psi(\{n_{jk}\})\rangle)
&\geq&-\log\biggl(\frac{N!}{\mathrm{perm}(A)}\prod_{j,k}f_{jk}^{n_{jk}}\biggr);
\end{eqnarray*}
the bound is saturated if and only if  there exists a pure state
$|\varphi\rangle$ such that
\begin{eqnarray}\label{eq:mubcond1}
|\langle\varphi|e^j_k\rangle|^2=f_{jk}\quad \forall j,k.
\end{eqnarray}
In that case, the GM of
$\bigl|\Psi\bigl(\{n_{jk}\}\bigr)\bigr\rangle$ and  any pure
symmetric $N$-partite state is additive .

In general, it is not easy to determine whether such a state exists
or not; the structure of the state space plays a crucial role here.
Here we are content to point out a connection with MUBs. Suppose the
$b$ bases $|e^j_k\rangle$ are mutually unbiased, and $n_{jk}=n_{j},
f_{jk}=1/d\; \forall j,k$. Then satisfying the set of constraints in
Eq.~(\ref{eq:mubcond1}) amounts to the existence of a pure state
that is mutually unbiased with all states $|e^j_k\rangle$.

\bt
   \label{thm:mubexistence}
   Suppose $\ket{e^j_k}$ for $j=0,1,\ldots,b-1 (b \le d+1)$
   and $k=0,1,\ldots,d-1$ are $b$
   MUBs, i.e.,
   $\abs{\braket{e^j_k}{e^l_m}}^2 = \frac1d(1-\d_{j,l})+\d_{j,l}
   \d_{k,m}$.  Then the $N$-partite $(N=d \sum^b_{j=1} n_j,~n_j\ge1)$ symmetric state
   $\ket{\Psi(d,\{n_j\})}
   := \sqrt{N!/\mathrm{perm}(A)}P_\mathrm{sym}
   \bigl(\bigotimes^b_{j=1} \bigotimes^d_{k=1} \ket{e^j_k}^{\otimes n_{j}}\bigr)$
   has GM lower bounded by
\begin{eqnarray}\label{eq:MUBGm}
   G(\ket{\Psi(d,\{n_j\})})\geq
   -\log\frac{N!}{d^N\mathrm{perm}(A)};
\end{eqnarray}
the bound is saturated if and only if  there exists a pure state
that is mutually unbiased to all states $\ket{e^j_k}$. If such a
state exists, then  the GM of $\ket{\Psi(d,\{n_j\})}$ and any pure
symmetric $N$-partite state $\ket{\Phi}$ is additive; that is,
\begin{eqnarray}
\label{eq:MUBGmAdd}
  G(\ket{\Psi(d,\{n_j\})} \ox \ket{\Phi})
   &=&G(\ket{\Psi(d,\{n_j\})})
  +
   G(\ket{\Phi}).  \nonumber\\
\end{eqnarray}
\et

For example, the inequality in Eqs.~(\ref{eq:MUBGm}) is saturated
and (\ref{eq:MUBGmAdd}) is  applicable when $b=2$ and the
$|e^{0,1}_k\rangle$'s are the eigenbases of $Z$ and $X$,
respectively, where $Z$ and  $X$ are phase operator and cyclic shift
operator, respectively. They  are defined according to their action
on the computational basis $|e_k\rangle=|e^0_k\rangle$,
\begin{eqnarray}\label{eq:XZ}
Z|e_k\rangle&=&\omega^k|e_k\rangle, \quad \omega=\mathrm{e}^{2\pi\mathrm{i}/d}\nonumber\\
X|e_k\rangle&=&|e_{(k+1)\, {\rm mod}\, d}\rangle,
\end{eqnarray}
where we use ``mod'' to denote the modulo operation. Since the
eigenbases of $X, Z$ and $XZ$ are mutually unbiased \cite{DEBZ10},
the $N$-fold tensor product of any eigenstate of $XZ$ is a closest
product state to $\ket{\Psi(d,\{n_j\})}$.

If $d$ is a prime power, there exists a complete set of $d+1$ MUBs
\cite{WF89,DEBZ10}. The inequality in Eqs.~(\ref{eq:MUBGm}) is
saturated and (\ref{eq:MUBGmAdd}) is  applicable if
$|e^j_k\rangle$'s are chosen from $b$ bases with $1\leq b\leq d$
from the complete set. However,  this is not the case if $b=d+1$,
since there is no pure state that is mutually unbiased to all states
in a complete set of MUBs \cite{WF89}.

\section{\label{sec:SICGM} symmetric states generated from  SIC~POVMs}

In a $d$-dimensional Hilbert space, a SIC~POVM \cite{Zau99, RBSC04,
App05,SG10} consists of $d^2$ outcomes that are  subnormalized
projectors onto pure states
$\Pi_j=\frac{1}{d}|\psi_j\rangle\langle\psi_j|$ for
$j=1,\ldots,d^2$, such that
\begin{eqnarray}\label{eq:SIC}
|\langle\psi_j|\psi_k\rangle|^2=\frac{1+d\delta_{jk}}{d+1}.
\end{eqnarray}
The condition  $\sum_j\Pi_j=I$ is already implied by the above
equation and need not  be imposed separately. Most known SIC~POVMs
are generated from a \emph{fiducial state} under the action of the
Heisenberg--Weyl (HW) group, which is generated by the two operators
$X,Z$ defined in Eq.~(\ref{eq:XZ}). A fiducial state $|\psi\rangle$
of the HW group obeys the following equations,
\begin{eqnarray}
|\langle\psi|X^{k_1}Z^{k_2}|\psi\rangle|=\frac{1}{\sqrt{d+1}}\quad
\end{eqnarray}
for all $(k_1,k_2)\neq (0,0)\mod d$. If $|\psi\rangle$ is a fiducial
state, then the $d^2$ states $X^{k_1}Z^{k_2}|\psi\rangle$ for
$k_1,k_2=0,1,\ldots,d-1$ form a SIC~POVM that is covariant with
respect to the HW group. The Clifford group is the normalizer of the
HW group that consists of unitary operators. Likewise, the extended
Clifford group is the larger group that contains also anti-unitary
operators. For any operator $U$ in the extended Clifford group,
$U|\psi\rangle$ is a fiducial state whenever $|\psi\rangle$ is.
Fiducial states and SIC~POVMs form disjoint orbits under the action
of the extended Clifford group. SIC~POVMs on the same orbit of the
extended Clifford group are \emph{equivalent} in the sense that they
can be transformed into each other with unitary or antiunitary
operations \cite{App05,Zhu10}.

In this section we study the GM of symmetric states generated from
SIC~POVMs,
 \be
   \ket{\Ps_d^{\mathrm{SIC}}} :=\sqrt{\frac{d^2!}{\mathrm{perm}(A)}}
   P_\mathrm{sym}(\ket{\ps_1}\ox\cdots\ox\ket{\ps_{d^2}}).
 \ee
Since the completely mixed  state is the only state that satisfies
$\langle\psi_j|\rho|\psi_j\rangle =1/d$ for $j=1, 2, \ldots, d^2$,
the GM of $\ket{\Ps_d^{\mathrm{SIC}}}$ cannot be computed according
to Theorem~\ref{thm:POVMstateGM}. In Appendix C, we derive the GM of
$\ket{\Ps_d^{\mathrm{SIC}}}$ by virtue of the special properties of
SIC~POVMs; the result is  summarized in the following theorem. \bt
   \label{thm:SIC-GM}
   Suppose $d^2$ normalized states $|\psi_j\rangle$ for $j=1,2,\ldots, d^2$ in a
   $d$-dimensional Hilbert space  satisfy $\sum^{d^2}_{j=1}\proj{\ps_j} = d I$, and $A$ is the Gram matrix of the $|\psi_j\rangle$'s; define
   $\ket{\Ps} = \sqrt{d^2!/\mathrm{perm}(A)}P_\mathrm{sym}(\ket{\ps_1}\ox\cdots\ox\ket{\ps_{d^2}})$. Then
   the following four statements are equivalent.
\begin{enumerate}
\item
   \label{item:sic}
   The $d^2$ states  $|\psi_j\rangle$ form a SIC~POVM, that is, they satisfy Eq.~(\ref{eq:SIC});

\item
   \label{item:closeproductstate}
   Up to global phases, the $d^2$ states $\ket{\ps_j}^{\ox d^2}$ for $j=1, \cdots,
   d^2$ are the only closest product states to
   $\ket{\Ps}$; and
\be
   \L^2(\ket{\Ps})= {d^2! \over {(d+1)^{d^2-1}}\mathrm{perm}(A)};
\ee

\item
   \label{item:innerproduct}
  The  $d^2$  states
   $\ket{\ps_j}^{\ox d^2}$ for $j=1,\cdots,d^2$ satisfy
\be
  \abs{ \bra{\ps_j}^{\ox d^2} \ket{\Ps} }^2
    ={d^2! \over {(d+1)^{d^2-1}}\mathrm{perm}(A)};
\ee
\item
   \label{item:closestSICPOVM}
  There exists a SIC~POVM consisting of  $d^2$ states $|\varphi_j\rangle$ for $j=1,2,\ldots,d^2$ such that
  $|\varphi_j\rangle^{\otimes d^2}$ satisfies
\be
   \abs{ \bra{\ph_j}^{\ox d^2} \ket{\Ps} }^2= {d^2! \over {(d+1)^{d^2-1}}\mathrm{perm}(A)}.
\ee
\end{enumerate}
\et

In dimension two, there is only one orbit of fiducial states, one of
them is given by \cite{Zau99,RBSC04,App05}
\begin{eqnarray}
\bigl|\psi^\mathrm{f}_2\bigr\rangle=\sqrt{(3+\sqrt{3})/6}\,|e_0\rangle+e^{\mathrm{i}\pi/4}\sqrt{(3-\sqrt{3})/6}\,|e_1\rangle.
\end{eqnarray}
Moreover, all SIC~POVMs are equivalent to the one thus generated.
The four states of each SIC~POVM form a regular tetrahedron when
represented on the Bloch sphere. Calculation shows that
$\mathrm{perm}(A)=\frac{8}{3}$, which, together with
Theorem~\ref{thm:SIC-GM}, implies that
\begin{eqnarray}
\L^2\bigl(\bigl|\Ps_2^{\mathrm{SIC}}\bigr\rangle\bigr)=\frac{1}{3},
\quad G\bigl(\bigl|\Ps_2^{\mathrm{SIC}}\bigr\rangle\bigr)=\log 3.
\end{eqnarray}
This result coincides with that obtained  by Martin et al.
\cite{MGBBB10} and Aulbach et al. \cite{AMM10}. They also showed by
numerics that the state generated from the SIC~POVM is the maximally
entangled pure symmetric four-qubit  state with respect to GM.

On the other hand, there is a one-parameter family of  fiducial
states in dimension three \cite{Zau99, RBSC04, App05,Zhu10},
\begin{eqnarray}
\label{eq:fiducial3}
 \bigl|\psi_3^\mathrm{f}(t)\bigr\rangle=\frac{1}{\sqrt{2}}\bigl(|e_1\rangle
   - \mathrm{e}^{\mathrm{i} t}|e_2\rangle\bigr);
\end{eqnarray}
and there is a one-to-one correspondence between orbits of the
extended Clifford group and the parameter $t$ for
$t\in[0,\frac{\pi}{3}]$. There are three kinds of orbits, two
exceptional orbits corresponding to the endpoints $t=0$ and
$t=\frac{\pi}{3}$, respectively, and infinitely many generic orbits
corresponding to $0<t<\frac{\pi}{3}$ . In addition, there is a
one-to-one correspondence between inequivalent SIC~POVMs and the
parameter $t$ for $0\leq t\leq \frac{\pi}{9}$; the SIC~POVMs on the
three orbits $t, \frac{2\pi}{9}-t, \frac{2\pi}{9}+t$, respectively,
are equivalent under unitary transformations. All these inequivalent
SIC~POVMs can be classified  in terms of  geometric phases
associated with fiducial states  \cite{Zhu10}. Here we provide an
alternative characterization in terms of GM.

\begin{figure}
  \includegraphics[width=7cm]{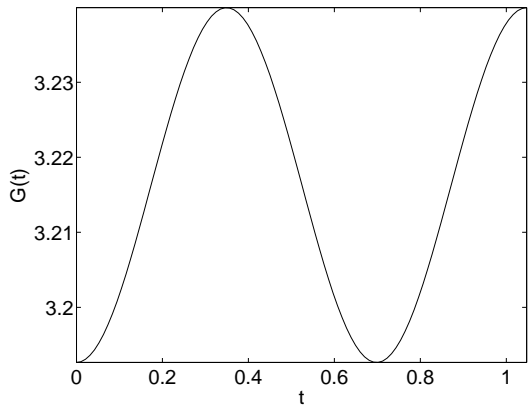}\\
  \caption{\label{fig:SIC3dGM} The GM of symmetric states $\bigl|\Ps_3^{\mathrm{SIC}}(t)\bigr\rangle$
   for $0\leq t\leq \frac{\pi}{3}$ generated from
  HW covariant SIC~POVMs in three-dimensional Hilbert space.
     Each SIC~POVM is uniquely specified by $G(t)$ up to the equivalence class; recall that
     there is a one-to-one correspondence between inequivalent SIC~POVMs and the parameter $t$ for $0\leq t\leq \frac{\pi}{9}$ \cite{Zhu10}. }
\end{figure}

Let $\bigl|\Ps_3^{\mathrm{SIC}}(t)\bigr\rangle$ denote the symmetric
state related to the SIC~POVM generated from the fiducial state
$\bigl|\psi_3^\mathrm{f}(t)\bigr\rangle$ under  the HW group, $A(t)$
the corresponding Gram matrix. Surprisingly, the permanent of $A(t)$
can be given by a simple formula:
$\mathrm{perm}(A(t))=\frac{27}{32}\bigl[61-\cos(9t)\bigr]$.
According to Theorem~\ref{thm:SIC-GM} and Eq.~(\ref{eq:GM}),
\begin{eqnarray}
G\bigl(\bigl|\Ps_3^{\mathrm{SIC}}(t)\bigr\rangle\bigr)&=&\log\frac{16[61-\cos(9t)]}{105}.
\end{eqnarray}
Figure~\ref{fig:SIC3dGM} shows  $G(t):=
G\bigl(\bigl|\Ps_3^{\mathrm{SIC}}(t)\bigr\rangle\bigr)$ for $0\leq
t\leq \frac{\pi}{3}$. Up to the equivalence class, each SIC~POVM is
uniquely specified by
$G\bigl(\bigl|\Ps_3^{\mathrm{SIC}}(t)\bigr\rangle\bigr)$. There is
only a small difference among the GM of symmetric states generated
from different SIC~POVMs. The minimum and the maximum of the GM are
obtained at the exceptional orbits $t=0$ and $t=\frac{\pi}{3}$,
respectively. Note that the SIC~POVMs on the orbits
$t=\frac{2\pi}{9}$ and $t=\frac{\pi}{9}$ are equivalent to the ones
on the orbits $t=0$ and $t=\frac{\pi}{3}$, respectively
\cite{Zhu10}.

\section{\label{sec:creation} Creating symmetric multi-qubit states}
In this section, we describe how to create symmetric multi-qubit
states mentioned in this article, following the approach by Bastin
et al.~\cite{Bas09}. For non-qubit systems, a generalization could
be made, but we do not know whether suitable physical systems exist.

Imagine we have an atom or trapped ion initially in the excited
state $|e\rangle$, and it has two stable ground states, labeled by
$|0\rangle$ and $|1\rangle$, which are  connected to the excited
state via emitting right-circularly ($R$) and left-circularly ($L$)
polarized light, respectively. If the detection of the emitted
photon is $R$ ($L$) then we know the atom  is now in the state
$|0\rangle$ ($|1\rangle$). The photon polarization carries the
which-way information. If we place a polarizer in $(|R\rangle
+|L\rangle)/\sqrt{2}$ to erase the which-way information and if the
detector behind the polarizer clicks, the atom is then projected to
an equal superposition of $|0\rangle$ and $|1\rangle$. In general,
if the polarizer is in $\alpha|R\rangle + \beta|L\rangle$, after
detecting the photon, the atom is in the state $\alpha|0\rangle +
\beta|1\rangle$.

Now suppose we have $N$ excited atoms and $N$ photon detectors. A
simultaneous detection of $N$ photons will project the $N$-atom
state, depending on the polarizers' settings. Suppose the $k$-th
polarizer is in $\alpha_k|R\rangle +\beta_k|L\rangle$. If the $k$-th
polarizer and detector sit close to the $k$-th atom and detect the
emitted photon by the nearby atom, and if  each detector-atom system
is far away from each other, then after detection of a photon at
every detector, the $N$-atom state is $\otimes_k (\alpha_k
|0\rangle_k +\beta_k |1\rangle_k)$. Bastin et al.~\cite{Bas09}
proposed the setup such that the $N$ detectors are placed in the
far-field, roughly at an equal distance to all atoms. Because of
multipath quantum interference (which-path information is erased),
after the detection, the atoms should be in a state that is
invariant under permutations, namely,
\begin{equation}
|\Psi\rangle \propto P_{\rm sym}\bigotimes_k (\alpha_k |0\rangle_k
+\beta_k |1\rangle_k).
\end{equation}
Thus, using the scheme by Bastin et al.~\cite{Bas09}, the symmetric
multi-qubit states discussed in this article can be realized in
principle, not just mathematical objects.

For non-qubit systems, one needs to first identify a physical system
that contains an excited state coupled to $d$ longlived sublevels
and that the decay into each level is associated with a
distinguishable which-way information. Deleting the which-way
information from the decay and the which-atom information will
enable the creation of symmetric multiqudit states considered in
this article. But we have not yet identified such a system.

Next, we discuss a possible measurement of the permanent of
$A_{ij}=\langle \psi_i|\psi_j\rangle$. Suppose one can prepare the
following symmetric $N$-particle  state
\begin{equation}
|\Psi\rangle= c P_{\rm sym} \mathop{\bigotimes}_{j=1}^N |\psi_j\rangle,
\end{equation}
with the normalization $c=\sqrt{N!/{\rm perm}(A)}$. If one chooses a
von Neumann measurement at each particle in the basis consisting of
$|\ph\rangle$ and the remaining orthonormal states, the simultaneous
detection of outcome $|\ph\rangle$ at all sites has the probability
\begin{equation}
\lambda(\ph)^2=\frac{N!}{{\rm perm}(A)} \prod_{j=1}^N
|\langle\ph|\psi_j\rangle|^2.
\end{equation}
Since  all $|\langle\ph|\psi_j\rangle|$'s are known and $\lambda(\ph)^2$ is
obtained from the measurement statistics, ${{\rm perm}(A)}$ can be inferred.
Ideally, one prefers $\lambda(\ph)^2$ to be as large as possible, as it
represents the probability of obtaining the desired outcome. One can also vary
$|\ph\rangle$; the probability $\lambda(\ph)^2$ is maximized when
$|\ph\rangle^{\otimes N}$ is the closest product state that results in the
geometric measure. According to a recent work by Martin et al.~\cite{MGBBB10},
the maximal overlap $\Lambda^2$ for highly entangled symmetric states decays
only inversely proportional to the number of qubits, so the proposed scheme
seems to be feasible even for medium size of systems that have been achieved
in various physical implementations.
\section{\label{sec:summary} summary}

We have studied the GM of pure symmetric states related to rank-one
POVMs and established its  connection with the maximum likelihood
principle in quantum state estimation theory. Based on this
connection, we provided a method for computing the GM of these
states and demonstrated its additivity property under certain
conditions. In particular, we proved the additivity of the GM of
pure symmetric multiqubit states whose Majorana points are
distributed within a half sphere, including all pure symmetric
three-qubit states. We then introduced a family of symmetric states
that are generated from MUBs and derived an analytical formula for
their GM. We also derived the GM of symmetric states generated from
SIC~POVMs and used it to characterize all inequivalent HW covariant
SIC~POVMs in three-dimensional Hilbert space. A scheme for creating
the symmetric multiqubit states studied in this article was also
proposed. Our studies promise a broad perspective of integrating two
important research areas in quantum information science, namely,
entanglement characterization and quantum state estimation.

\section*{Acknowledgment}
We thank Markus Grassl for discussions on the symmetric states
generated from SIC~POVMs and for his valuable  comments on the
manuscript. We also thank Daniel Greenberger for discussions on the
connection between the GM of symmetric states and the maximum
likelihood principle. H.Z. thanks Berthold-Georg Englert, Yong Siah
Teo and  Hui Khoon Ng for discussions on the maximum likelihood
methods. The Centre for Quantum Technologies is funded by the
Singapore Ministry of Education and the National Research Foundation
as part of the Research Centres of Excellence programme. T.-C.W.
acknowledges support from NSERC and MITACS.

\section*{Appendix A: Additivity of the GM of $|\Psi(\{n_{j}\})\rangle$}
In this appendix, we prove Eq.~(\ref{eq:POVMstateGMadd}) when the
maximum of the likelihood functional $\mathcal{L}(\rho)$ can be
obtained at a pure state. Equation~(\ref{eq:POVMstateGMadd}) is
equivalent to
 \be
 \label{appendixAeq:add}
\L^2(|\Psi(\{n_{j}\})\rangle \ox \ket{\Ph}) =
\L^2(|\Psi(\{n_{j}\})\rangle)\L^2(\ket{\Ph}).
 \ee
According  to the definition, the l.h.s is never smaller than the
r.h.s; so it suffices to show that $\L^2(|\Psi(\{n_{j}\})\rangle \ox
\ket{\Ph})\leq\L^2(|\Psi(\{n_{j}\})\rangle)\L^2(\ket{\Ph})$. Suppose
$|\Psi(\{n_{j}\})\rangle$ is shared over the parties $A_1,
A_2,\ldots, A_N$, and $|\Phi\rangle$ over the parties $B_1,
B_2,\ldots, B_N$; suppose $|a_k\rangle$ is the $k$-th member of the
multiset consisting of $n_1$ copies of $|\psi_1\rangle$, $n_2$
copies of $|\psi_2\rangle$ and so on. Then according to
Proposition~\ref{pro:symmetric},
 \bea
     &&
    \L^2(|\Psi(\{n_{j}\})\rangle \ox \ket{\Ph})\nonumber\\
  &=&\max_{\ket{\ph}} \bigl| \bra{\ph}^{\ox N}|\bigl(|\Psi(\{n_{j}\})\rangle
\ox \ket{\Ph}\bigr) \bigr|^2
    \nonumber\\
    &=& \frac{N!}{\mathrm{perm}(A)}~\max_{\ket{\ph}}
    \bigg| \biggl(\bigotimes^{N}_{k=1} \bra{\ph}_{A_k,B_k}\ket{a_k}_{A_k}\biggr)
     \ket{\Ph}_{B_1,\ldots,B_{N}} \bigg|^2
    \nonumber\\
    &=& \frac{N!}{\mathrm{perm}(A)}~
    \max_{\ket{\ph}}
    \Bigg[\biggl(
    \bigotimes^{N}_{k=1}
    \bigl|\bra{\ph}_{A_k,B_k}\ket{a_k}_{A_k}\bigr|^2
    \bigg)\nonumber\\
    &&\times
    \biggl|\bigg(
    \bigotimes^{N}_{k=1}
    {\bra{\ph}_{A_k,B_k}\ket{a_k}_{A_k}
    \over\bigl|\bra{\ph}_{A_k,B_k}\ket{a_k}_{A_k}\bigr|}
    \bigg)
    \ket{\Ph}_{B_1,\ldots,B_{N}}
    \bigg|^2\Biggr]
    \nonumber\\
    &\leq&\L^2(\ket{\Ph}) \frac{N!}{\mathrm{perm}(A)}
    \max_{\r}
    \prod^{N}_{k=1}
    \big(
    \bra{a_k}_{A_k}\,\r_{A_k}\!\ket{a_k}_{A_k}
    \big)        \nonumber\\
    &=&\L^2(\ket{\Ph}) \frac{N!}{\mathrm{perm}(A)}
    \max_{\r}\mathcal{L}(\rho)\nonumber\\
    &=& \L^2(|\Psi(\{n_{j}\})\rangle)\L^2(\ket{\Ph}).
 \eea
Here the last equality is due to our assumption that the maximum of
the likelihood functional $\mathcal{L}(\rho)$ can be obtained at a
pure state.

\section*{Appendix B: perm(A) in Eq.~(\ref{eq:gDicke})}

Note that the entries of the Gram matrix $A$ in
Eq.~(\ref{eq:gDicke}) only take on five different values
$0,1,\sin(\theta/2), \pm\cos(\theta/2)$. In contrast with the
computation of the permanent of a generic matrix,  the permanent of
$A$ can be computed efficiently as follows:
\begin{eqnarray}
&&\mathrm{perm}(A)= \Biggl(\prod_{j,k=0}^1
n_{jk}!\Biggr)\sum_{\{a,b,c,f,g\}}(-1)^{g-a}\Bigl(\cos\frac{\theta}{2}\Bigr)^{2f+2g}\nonumber\\
&&\times\Bigl(\sin\frac{\theta}{2}\Bigr)^{2a+2b+2c-2g}{n_{00}\choose
a,
b, n_{00}-a-b}\nonumber\\
&&\times{n_{01}\choose c,f,n_{01}-c-f}{n_{10}\choose g, a+c-g,
n_{10}-a-c}\nonumber\\
&&\times{n_{11}\choose a+b-g, f+g-a, n_{11}-b-f},
\end{eqnarray}
where the summation is restricted to the set of nonnegative integers
$\{a,b,c,f,g\}$ satisfying the following constraints,
\begin{eqnarray}
&&a+b\leq n_{00},\;\; c+f\leq n_{01},\;\; a+c\leq
n_{10},\;\; b+f\leq n_{11},\nonumber\\
&&g\leq a+b, \quad g\leq a+c,\quad f+g\geq a.
\end{eqnarray}

\section*{Appendix C: Proof of Theorem~\ref{thm:SIC-GM}}

   We begin by proving the implication \ref{item:sic}$\Ra$\ref{item:closeproductstate}.
   According to Proposition~\ref{pro:symmetric} and the definition of
   a SIC~POVM,
\bea
    \L^2(\ket{\Ps})&=&
    \L^2(\ket{\Ps_d^{\mathrm{SIC}}})
    = \max_{\ket{\ph}} \bigl| \bra{\ph}^{\ox d^2} \ket{\Ps_d^{\mathrm{SIC}}}
    \bigr|^2\nonumber\\
    &=&\frac{d^2!}{\mathrm{perm}(A)}~\max_{\ket{\ph}} \prod^{d^2}_{j=1} \abs{ \braket{\ph}{\ps_j}
    }^2.
\eea
  Here the maximization is subjected to the completeness condition $\sum_j \abs{ \braket{\ph}{\ps_j} }^2 =
  d$ and the following condition
\bea
    \label{ea:symmetricoperator}
    \sum_j \abs{ \braket{\ph}{\ps_j} }^4
    =  \frac{2d}{d+1},
\eea since a SIC~POVM is a 2-design \cite{RBSC04}: $\sum_j
|\psi_j\rangle\langle\psi_j|\otimes|\psi_j\rangle\langle\psi_j|=[2d/(d+1)]\Pi_{\rm
sym}$, where $\Pi_{\rm sym}$ is the projector onto the bipartite
symmetric subspace. Hence,

\bea
    \label{eqa:maximumSICPOVM}
    \L^2(\ket{\Ps_d^{\mathrm{SIC}}})
    &\leq& \frac{d^2!}{\mathrm{perm}(A)}~\max_{0\leq p_j\leq 1,\; \sum_j p_j = d, \atop\sum_j p_j^2 = \frac{2d}{d+1}}
    \prod^{d^2}_{j=1} p_j\nonumber\\
    &=&{d^2! \over (d+1)^{d^2-1}\mathrm{perm}(A)}.
\eea
  The maximum in the above equation is obtained if all $p_j$'s are
  equal to $1/(d+1)$ except one of them, which is equal to 1. A state can  satisfy these conditions if and only it  belongs to the
  SIC~POVM, hence, the implication \ref{item:sic} $\Ra$
\ref{item:closeproductstate} follows.

The implication  \ref{item:closeproductstate}
$\Ra$\ref{item:innerproduct} is obvious.  The implication
\ref{item:innerproduct}$\Ra$\ref{item:sic} and
\ref{item:closestSICPOVM} can be shown as follows, \bea
  \label{ea:innerproduct}
    \abs{ \bra{\ps_j}^{\ox d^2} \ket{\Ps} }^2
       &=&\frac{d^2!}{\mathrm{perm}(A)} \prod^{d^2}_{k=1} \abs{ \braket{\ps_j}{\ps_k} }^2
    \nonumber\\
    &\leq&\frac{d^2!}{\mathrm{perm}(A)}
    \bigg(
    {{\sum^{d^2}_{k=1,k\ne j} \abs{ \braket{\ps_j}{\ps_k} }^2 }
    \over
    {d^2-1}}
    \bigg)^{d^2-1}
    \nonumber\\
    &=&{d^2! \over {(d+1)^{d^2-1}}\mathrm{perm}(A)}.
\eea
 The inequality is saturated  if and only if $|\langle\psi_j|\psi_k\rangle|^2=1/(d+1),~\forall j,k$ and $j\neq
 k$; that is, the $\ket{\ps_j}$'s  form a SIC
  POVM, which implies \ref{item:sic} and \ref{item:closestSICPOVM}.

It remains to show the implication
\ref{item:closestSICPOVM}$\Ra$\ref{item:sic},
 \bea
|\langle\varphi_j|^{\otimes d^2}\Psi\rangle|^2&=&
    \frac{d^2!}{\mathrm{perm}(A)} \prod^{d^2}_{k=1} \abs{ \braket{\ph_j}{\ps_k} }^2
    \nonumber\\
    &=&
    \frac{d^2!}{\mathrm{perm}(A)}
    \prod^{d^2}_{k=1}
    \bigg(
    \prod^{d^2}_{j=1} \abs{ \braket{\ph_j}{\ps_k} }^2
    \bigg)
    ^{\frac{1}{d^2}}
    \nonumber\\
    &\le&
    \frac{d^2!}{\mathrm{perm}(A)}
    \max_{|\ps\rangle}
    \prod^{d^2}_{j=1}
    \abs{ \braket{\ph_j}{\ps} }^2
    \nonumber\\
    &\leq&
    {d^2! \over {(d+1)^{d^2-1}}\mathrm{perm}(A)}.
\eea
  Here the second inequality follows from Eq.~(\ref{eqa:maximumSICPOVM}), recall that the $|\ph_j\rangle$'s form a SIC
  POVM.  The inequalities are saturated if and only if  the $\ket{\ps_j}$'s are
 the same as the  $\ket{\ph_j}$'s up to some permutation and phase factors;  which implies 1. 

\end{document}